*Electron paramagnetic resonance signature of point defects in neutron irradiated hexagonal boron nitride*

J.R. Toledo[1], D.B. de Jesus[1], M. Kiania[2], A.S. Leal[3], C. Fantini[1], L.A. Cury[1], G. A.M. Sáfar[1], I. Aharonovich[2], K. Krambrock[1]

[1]*Departamento de Física, Universidade Federal de Minas Gerais (UFMG), Belo Horizonte, MG, Brazil*

[2]*School of Mathematical and Physical Sciences, University of Technology, Sydney, Australia*

[3]*Centro de Desenvolvimento da Tecnologia Nuclear (CDTN), Belo Horizonte, MG, Brazil*

*\* Corresponding author: klaus@fisica.ufmg.br*

**Abstract**

Hexagonal boron nitride (h-BN) is an attractive van der Waals material for studying fluorescent defects due to its large bandgap. In this work, we demonstrate enhanced pink color due to neutron irradiation and perform electron paramagnetic resonance (EPR) measurements. The new point defects are tentatively assigned to doubly-occupied nitrogen vacancies with ($S = 1$) and a zero-field splitting (D = 1.2 GHz). These defects are associated with a broad visible optical absorption band and near infrared photoluminescence band centered at ~ 490 nm and 820 nm, respectively. The EPR signal intensities are strongly affected by thermal treatments in temperature range between 600 to 800ºC, where also the irradiation-induced pink color is lost. Our results are important for understanding of point defects in h-BN and their deployment for quantum and integrated photonic applications.

## 1. Introduction

Hexagonal boron nitride (h-BN) is a wide bandgap ( ~ 6 eV) two dimensional (2D) semiconductor with excellent thermal, mechanical and optical properties [1–4]. Recently, it has emerged as promising material for nanophotonics due to its ability to host single photon emitters as well as a subject of rigorous studies of its excitonic and phononic properties [5–10]. Amongst the various fluorescent defects in h-BN, there is a great interest to identify new paramagnetic centers that can be used in quantum information science, especially those with high spin because the polarization of recombination luminescence is dependent on the spin state of the center [11]. Such single defects exist in other materials [12], with most prominent example being the nitrogen vacancy in diamond [13–15].

In h-BN so far, the most investigated paramagnetic centers are the so-called one boron centers (OBC) and three boron centers (TBC), both related with single electron trapped ($S = 1/2$) nitrogen vacancies [16–19]. The OBC centers are observed in samples that underwent an oxidation, and the electron is interacting strongly with the nuclear spin of one of the nearest neighbor boron atoms of the nitrogen vacancy. The TBC complexes have the trapped electron of the nitrogen vacancy interacting with the nuclear spin of three nearest neighbor equivalent boron atoms. Moore et al. [18] noticed that the latter TBC center is found in samples with carbon impurities. Katzir et al. [16] stated that this carbon related TBC centers is a yellow color center, and involved in a blue luminescence at 3.1 eV. While the TBC centers have an energy level at about 1.1 eV below the conduction band, the energy level introduced by substitutional carbon impurities on is localized at about 4.2 eV. On the other hand, TBC centers can be also produced by high-energetic electrons in h-BN in the absence of carbon impurities [17]. Motivated by the presence of paramagnetic defects in h-BN and the recent works on isolated individual defects that can emit single photons, we performed a systematic study of h-BN material that undergoes neutron irradiation. Our results suggest that neutron irradiation induces new type of paramagnetic defects with optical absorption in the visible spectral range.

## 2. Experimental

Commercial h-BN ultrafine powder samples (Graphene supermarket) 99.0 % purity, particles size ~70 nm and specific surface area ~20 m²/g were neutron irradiated in nuclear reactor Triga Mark I IPR-R1 of CDTN, Brazil with thermal flux of 4 x $10^{12}$ n·cm$^{-2}$s$^{-1}$ for different times reaching integrated dose of approximately $3\times10^{17}$, $6\times10^{17}$, $12\times10^{17}$, $23\times10^{17}$ n·cm$^{-2}$. The powdered samples were neutron irradiated in Cd capsules for which thermal neutrons are blocked, and mostly energetic neutrons pass the samples. The identification of samples in relation to the integrated dose is shown in the table 1.

*Table 1-* *Labels to samples in relation to time and dose of neutron irradiation.*

| Sample | Time of neutron irradiation (h) | Integrated dose ($10^{17}$ n·cm$^{-2}$) |
|---|---|---|
| **0h (pristine)** | 0 | 0 |
| **2h** | 2 | 3 |
| **4h** | 4 | 6 |
| **8h** | 8 | 12 |
| **16h** | 16 | 23 |

Optical properties of the samples were investigated by photoluminescence (PL), reflectance and single-photon correlation measurements. Electron paramagnetic resonance (EPR) was employed to study the nature of the paramagnetic centers and isochronal annealing was performed to investigate the stability of defects. Photoluminescence (PL) measurements were done using a continuous laser in UVA region (375 nm, 8mW) or alternatively in the VIS region (532 nm, 10 mW) and the PL spectra were acquired with an ANDOR-Shamrock 303i spectrometer with a CCD detector. Optical reflectance spectra were measured with a Shimadzu UV3600 spectrometer in the region of 200 nm to 900 nm. The EPR spectra were recorded on a MiniScope MS 400 spectrometer (Magnettech, Germany) operating at X-band (9.43 GHz) using quartz tubes (Wilmad) as sample holder. Low temperatures measurements were measured by a He flux cryosystem (Oxford) in the 4.2 to 300 K range. Isochronal annealing experiments (15 min.) were performed in a tubular furnace (Lindberg) with controlled atmosphere in the temperature range from 500ºC to 800ºC. For single photon emission measurements, a scanning confocal microscope with Hunbury Brown and Twiss (HBT) configuration was employed. The samples were excited with a 532 nm laser using a high numerical aperture (NA = 0.9) objective and the signal was collected using same objective and directed into either a spectrometer or two avalanche photo detectors for photon statistics. A long pass filter (568 nm) and a dichroic mirror were used to filter the laser line. The measurements were performed at room temperature and the setup spatial resolution is ~ 400 nm.

## 3. Results and Discussion

### 3.1 Optical Measurements

As-received hexagonal boron nitride (h-BN) powders showed optical absorption (measured as reflectance spectra) only in the ultraviolet (UV), near bandgap region, which gives it a white color (Figure 1a). Neutron irradiation changed the color to pink which is correlated with irradiation dose (see photos in the inset of Figure 1a). The pink color is due to the creation of two absorption bands: (i) a broad UV absorption band centered near band gap and (ii) a broad absorption band in the visible spectral range centered at about 490 nm. These features result in the pink color for the neutron irradiated powders. The intensity of both bands increases with the dose of the irradiation, which indicates the formation of color centers, i.e. point defects that induce energy levels in the band gap of h-BN.

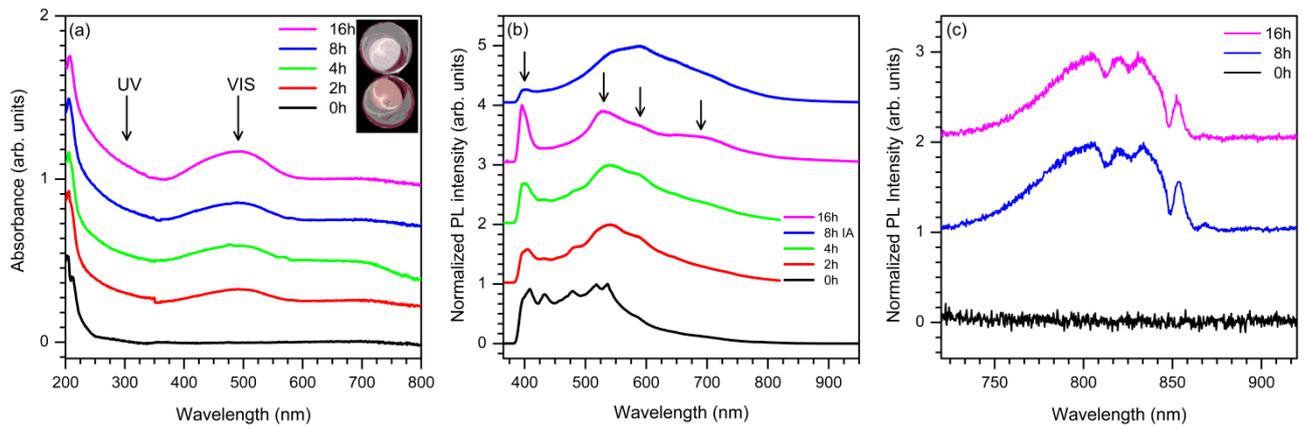

*Figure 1- (a)* Absorbance spectra in the UV-VIS spectral range for neutron irradiated h-BN powders (measured in reflectance mode). The inset shows the pristine (white) and sample 8h (pink), illustrating the pink color after neutron irradiation. *(b)* Photoluminescence spectra of non- and neutron irradiated h-BN powder samples (excited by UV laser 375 nm, 8mW) at room temperature. In legend, 1A refers to an isochronal annealing (800°C, 1h) made on sample 8h. The arrows indicate the positions of the bands that reduces after isochronal annealing (400 nm) and the bands which are created after neutron irradiation. *(c)* PL spectra for samples 0h, 8h and 16h excited at 532 nm showing near infrared luminescence.

Figure 1b shows room temperature PL measurements of the different samples measured under 375 nm excitation. In all cases the PL spectra are composed of broad emission bands and a narrower band at around 400 nm which intensifies with the dose of irradiation. This peak is also present in non-irradiated sample, but it is superimposed with other more intense peaks showing phonon-like structure with energy separation of 1389 cm$^{-1}$ corresponding to in-plane TO phonons [20]. It is important to note that neutron irradiation intensifies some optical transitions and reduces others. Comparing the samples 2h, 4h and 16h to the non-irradiated sample, it is possible to observe that the peak at about 400 nm increases with dose; new broad band emissions appear at about 530 nm, 590 nm and 690 nm, but on the other hand narrower peak emissions between 400 and 600 nm are lost with increasing dose. Isochronal annealing at 800°C for 1 hour (sample 8h - 1A), reduces strongly the 400 nm emission (Figure 1b), while the broad emission bands at 530 nm, 590 nm and 690 nm are still observed. This confirms that the sample did not return to its original state. PL measurements (Figure 1c) of irradiated samples excited by 532 nm compared with reference sample 0h show that neutron irradiation induces broad luminescence bands in the near infrared spectral region centered at about 820 nm.

### 3.3 Electron paramagnetic resonance

EPR measurements carried out at X-band frequencies and room temperature of as-received (non-irradiated, 0h) h-BN powders did not show any EPR signature. On the other hand, three different

EPR signals, which we name A, B and C were observed for the neutron irradiated samples (Figure 2a). The intensity of the EPR spectra of all paramagnetic defects increased with the applied irradiation dose. Powder spectra simulations using the Easyspin@ software reveal that defects A and B may be attributed to the OBC and TBC defects (Figure 2b) [16,17,21], respectively, as will be described later. However, to our knowledge the EPR spectrum labeled C is a yet unknown paramagnetic center (Figure 2c).

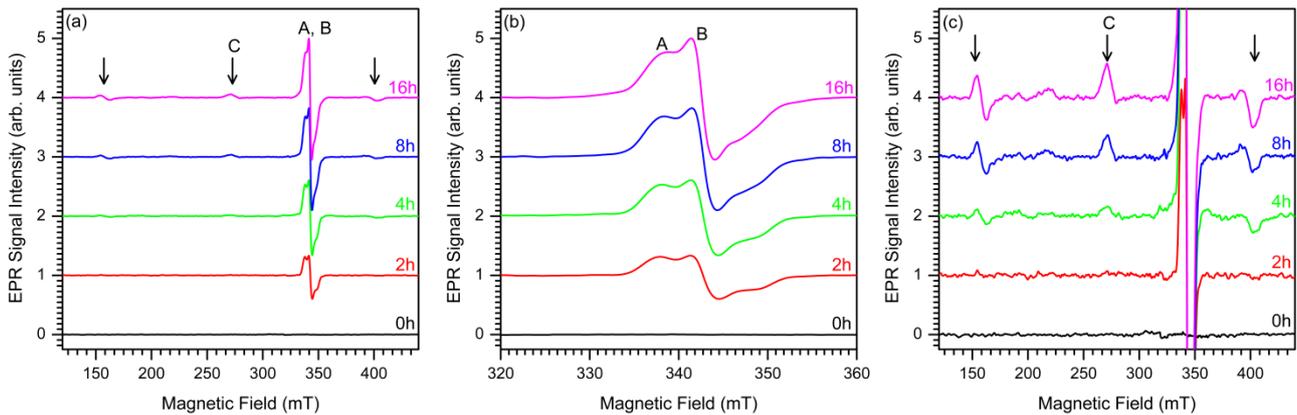

*Figure 2 -* *EPR spectra (9.43 GHz) of h-BN powder as a function of neutron irradiation dose measured at room temperature: (**a**) in the 120 to 440 mT range (**b**) zoomed in region from the 320 to 360 mT range (**c**) zoomed in region around "C" defect with lines indicated by arrows.*

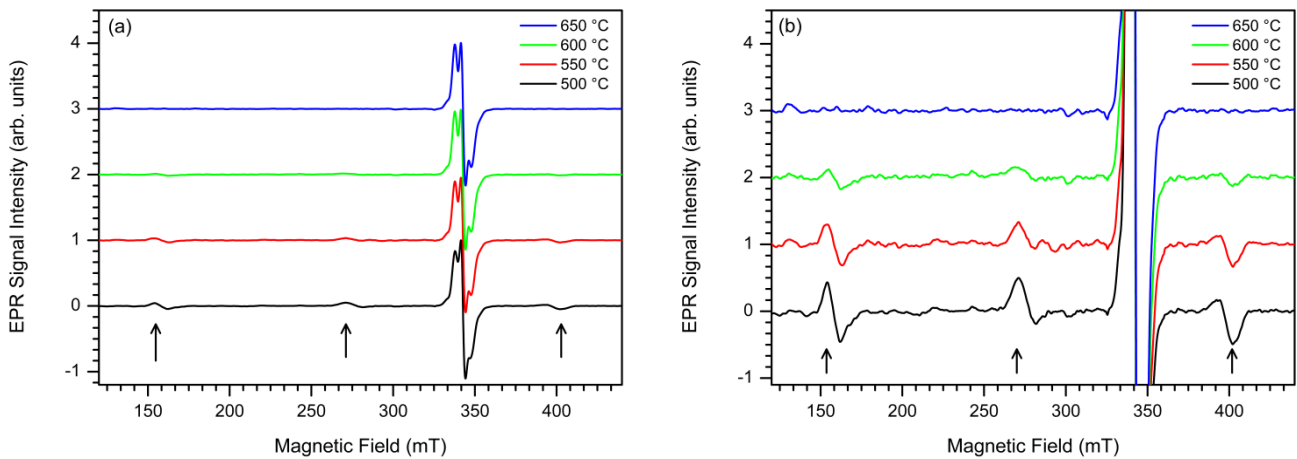

*Figure 3 -* *(**a**) EPR spectra (9.43 GHz) of h-BN powder for different isochronal (15 min.) annealing in between 500 and 800°C (15 min.) of the neutron irradiated sample (16h) with arrows indicating defect C and (**b**) zoomed in for visualization of lines associated with defect C.*

To understand the origin of the induced paramagnetic defects, we performed isochronal annealing experiments in Argon atmosphere on the neutron irradiated samples (15 min. at each temperature). Optical characterization of the samples after annealing shows that the pink color is lost for temperatures at about 600°C which also coincides with the loss of the signature of the EPR lines attributed to defect C (Figure 3a, marked by arrows). Therefore, it is likely to attribute the EPR lines C to the pink color center. The EPR lines associated to the A and B centers are stable under these annealing conditions, however, they decrease in intensity for annealing temperatures above 700ºC (not shown). In order to test the hypothesis that the defect C annealed out from the sample during the heat treatments, or simply is lost because of depopulation, gamma irradiation from a $^{60}$Co (0 - 400 kGy) was used for the sample 8h treated after hear treatment at 800ºC. It was found that after gamma irradiation the signal intensities of A and B were enhanced again (not shown), however, the pink color and the EPR signals due to C did not appear.

To analyze in more detail the microscopic origin of the three powder EPR lines (~150 mT, 280 mT and 410 mT) associated to defect C, three hypotheses are discussed: (i) the three lines belong to a center with three very different g values including strong deviation from the g value of the electron, (ii) the three lines are due to a hyperfine interaction with a nuclear spin of $I = 1$ or (iii) the lines are caused by electronic fine structure from a paramagnetic center with spin S ≥ 1. On a first sight, the defects C induced by neutron irradiation should be of intrinsic origin because they were observed in h-BN from different sources, not only in h-BN samples shown in this work but also in h-BN nanotubes produced by others [22]. Strong deviation of g factors from that of the electron is not expected for intrinsic defects in h-BN. Second, strong hyperfine interaction due to a nuclear spin of $I = 1$ of the order of ~1.5 GHz is also not very likely, because this is associated with strong localization of wavefunction around the nuclear spin, so this feature is not predicted [19]. Therefore, the best interpretation of the nature of defects C is that they belong to a center with high spin state of $S ≥ 1$. To confirm this, we performed EPR spectra calculations using the Easyspin$^@$ software (Figure 4) with a spin Hamiltonian consisting of the electron Zeeman interaction and the electronic fine structure interacting for a spin $S = 1$:

$$H = \mu_B \vec{B} \cdot \vec{\vec{g}} \cdot \vec{S} + \vec{S} \cdot \vec{\vec{D}} \cdot \vec{S} \qquad (1)$$

where the symbols μ$_B$, B, S, g and D represent respectively the Bohr magneton, the magnetic field, the electronic spin, the g and electronic fine structure tensors. The calculations show that defects C can be attributed to a center with high spin $S = 1$, g value of 2.0(1) and electronic fine structure interacting D of 1.2(1) GHz assuming Lorentzian line shape and linewidth of about 9(1) mT (Figure 4a). In this calculation the low-field EPR line (~150 mT) is due to the forbidden transition of type $\Delta m_S = \pm 2$ in this $S = 1$ system, the so called half-field transition. Low-temperature EPR measurements (not shown) are consistent with this interpretation because below 10 K the intensity of the forbidden transition increased much more compared with the allowed transitions. Finally, the central EPR lines A and B which we had associated to the OBC and TBC defects were satisfactorily reproduced by spectra calculations using published spin Hamiltonian parameters [16,17] (Figure 4b).

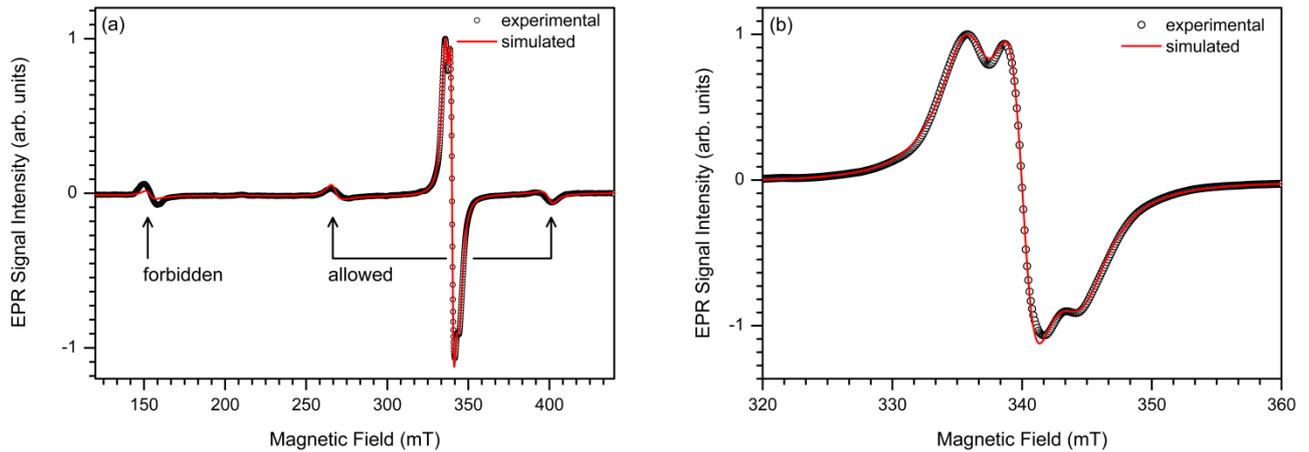

*Figure 4 - (a) Experimental EPR spectrum (4.2 K, 9.43 GHz) of sample 16h of h-BN powder (black) and calculated spectrum (red) for defects C assuming S = 1 center (parameters see text). The transition at about 150 mT is due to a forbidden transition ($\Delta m_S = \pm 2$). Forbidden and allowed transitions are indicated by arrows. (b) Experimental EPR spectrum (4.2 K, 9.43 GHz) of sample 16h of h-BN powder (black) and spectra calculation (red) for A = OBC and B = TBC defects.*

The most potential candidate for the spin ($S = 1$) defect is the doubly-occupied nitrogen vacancy-related defect which is associated to the pink color (broad absorption band centered at 490 nm in the visible spectral range) of neutron irradiated h-BN because experimentally, only nitrogen vacancy-related defects in h-BN have been identified [16–18,21]. From the g - factor (~ 2) of the pink color center, it is important to conclude that heavy atoms are not present or, at least, are not associated to the observed EPR spectrum. In addition, defects C were produced also by neutron irradiation in h-BN from other sources [22] indicating intrinsic nature of defect C, excluding extrinsic impurities. Theoretical calculations on point defects in h-BN are controversial. DFT calculations on isolated nitrogen vacancies suggest that the $V_N$ defects act as single donors with singlet ground state [23], while pair defects based on $V_N$-$C_B$, frequently discussed as potential candidate for single photon emission in h-BN, are predicted with singlet or triplet ground states [19,24,25]. More, nearest neighbor $V_N$-$N_B$ defects are also considered as possible candidate for single photon emission in h-BN [26].

On the other hand, the TBC centers are associated with the single electron trapped nitrogen vacancy, i.e. spin ($S = 1/2$), which originate a yellow color in h-BN due to broad absorption band in the UV spectral region. It is likely that at about 500ºC the nitrogen vacancy starts to diffuse and some may recombine with nitrogen interstitials produced also by the neutron irradiation as primary Frenkel pairs.

**3.2 Single photon emission**

Finally, we checked also the presence of single photon emitters in neutron irradiated h-BN samples. PL measurements from two irradiated samples (8h and 16h) are shown in Figure 5. Representative confocal maps of samples 8h and 16h are depicted in Figure 5a and 5b respectively. We found broad PL emission centered around 820 nm (as shown in Figure 1c) in most of the bright spots in the confocal map. However, single photon emitters could also be identified in both samples. Single photon emitters found in the confocal maps in Figure 5a and 5b are indicated by pink and blue arrows, respectively. PL spectrum of these single photons are shown in Figure 5c. Autocorrelation measurements with $g^{(2)}(\tau = 0)$ below 0.5 (Figure 5d) confirm the quantum nature of the emission from these defects. Note that we observed single photon emission from pristine (non-irradiated) samples, therefore based on the results here, we cannot confirm whether these emitters are created or activated during irradiation. In addition, our attempts at generating optically dependent magnetic resonance signatures at room temperature (under zero magnetic field) from these defects were not successful.

Notably, the defects that produce the absorption band that generate the pink color that is most likely associated with the $S = 1$ defect is excited at much higher energies (375 nm). We did not observe single emitters under these excitation conditions. However, it may be plausible that these defects have a different charge state, or have a significantly lower quantum efficiency. A similar effect is currently under debate in the literature for the neutral silicon vacancy in diamond that is EPR active [27,28] but has not been conclusively isolated on a single level [29]. Finally, our results can assist in explaining the recent results of isolated defects in h-BN [30] that exhibited reduced luminescence as a function of applied magnetic field.

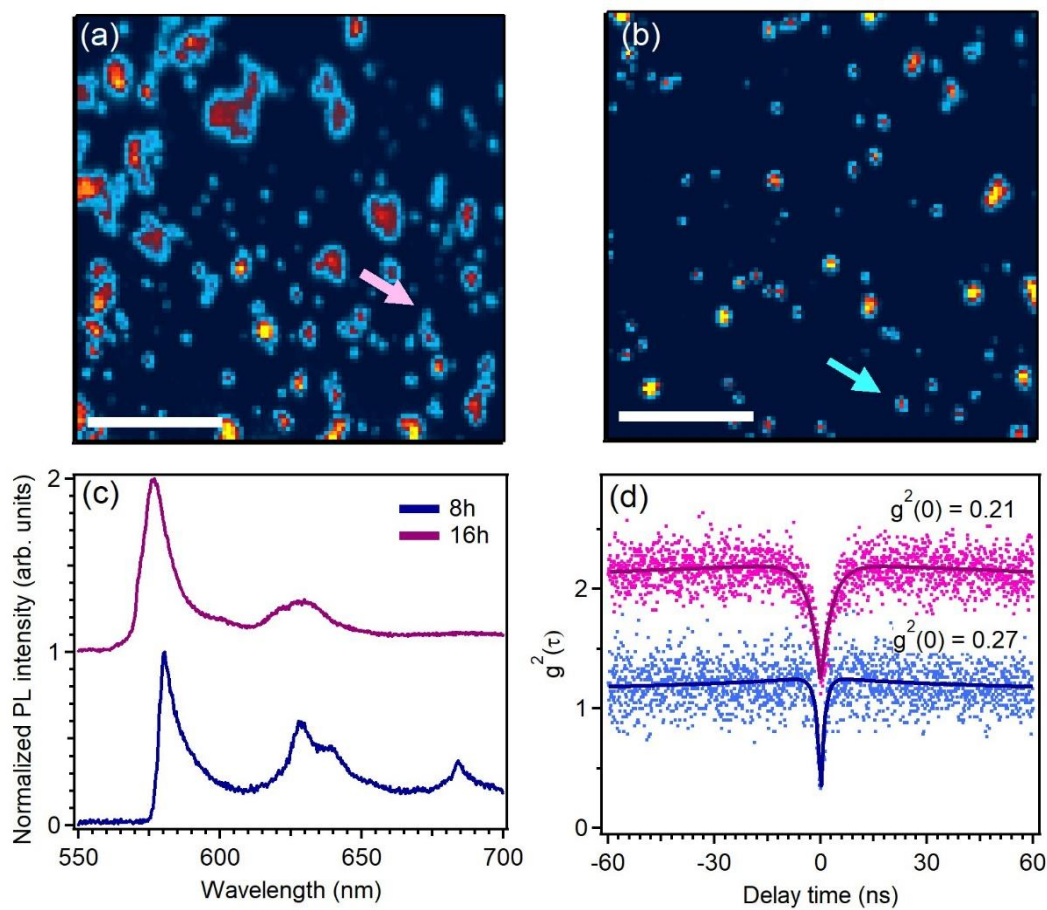

***Figure 5*** - *Photoluminescence measurement of the neutron irradiated h-BN under sub bandgap excitation (532 nm). Representative PL confocal maps **(a)** and **(b)** of samples 8h and 16h respectively. **(c)** PL spectra obtained from single emitters in 16h (top) and 8h (bottom) irradiated samples as indicated with arrows in the confocal maps. **(d)** Autocorrelation measurement from emissions shown in (c) confirming the quantum nature of the emission. Single photon emission was also found in the pristine h-BN sample (data not shown). Scale bars are 20 μm.*

## *Conclusions*

To conclude, we demonstrated that neutron irradiation in h-BN samples creates a previously unknown paramagnetic defect associated with an optical absorption band that gives a pink color to the materialand also a broad near infrared luminescence centered at 820 nm. EPR measurements were used to confirm that the irradiation-induced color center is composed by a state with high electronic spin ($S = 1$) and isochronal annealing experiments illustrate a behavior that can be associated to nitrogen vacancies that trap two electrons. These results are an important step towards identification of defects with high electronic spin, and further understanding of emission properties of h-BN.

**Acknowledgements**

This work was supported by Brazilians agencies CNPq, CAPES and FAPEMIG.Financial support from the Australian Research council (via DP180100077, LP170100150), the Asian Office of Aerospace Research and Development grant FA2386-17-1-4064, the Office of Naval Research Global under grant number N62909-18-1-2025 are gratefully acknowledged.

**References**

[1]   K. Watanabe, T. Taniguchi, and H. Kanda, Nat. Mater. **3**, 404 (2004).

[2]   J. H. Edgar, T. B. Hoffman, B. Clubine, M. Currie, X. Z. Du, J. Y. Lin, and H. X. Jiang, J. Cryst. Growth **403**, 110 (2014).

[3]   G. Cassabois, P. Valvin, and B. Gil, Nat. Photonics **10**, 262 (2016).

[4]   Q. Weng, D. G. Kvashnin, X. Wang, O. Cretu, Y. Yang, M. Zhou, C. Zhang, D. M. Tang, P. B. Sorokin, Y. Bando, and D. Golberg, Adv. Mater. **29**, 1 (2017).

[5]   L. Schué, I. Stenger, F. Fossard, A. Loiseau, and J. Barjon, 2D Mater. **4**, 1 (2017).

[6]   A. J. Giles, S. Dai, I. Vurgaftman, T. Hoffman, S. Liu, L. Lindsay, C. T. Ellis, N. Assefa, I. Chatzakis, T. L. Reinecke, and others, Nat. Mater. **17**, 134 (2018).

[7]   T. T. Tran, C. Zachreson, A. M. Berhane, K. Bray, R. G. Sandstrom, L. H. Li, T. Taniguchi, K. Watanabe, I. Aharonovich, and M. Toth, Phys. Rev. Appl. **5**, 034005 (2016).

[8]   M. Kianinia, C. Bradac, B. Sontheimer, F. Wang, T. T. Tran, M. Nguyen, S. Kim, Z.-Q. Xu, D. Jin, A. W. Schell, and others, Nat. Commun. **9**, 874 (2018).

[9]   R. Bourrellier, S. Meuret, A. Tararan, O. Stéphan, M. Kociak, L. H. G. Tizei, and A. Zobelli, Nano Lett. **16**, 4317 (2016).

[10]  N. R. Jungwirth, B. Calderon, Y. Ji, M. G. Spencer, M. E. Flatté, and G. D. Fuchs, Nano Lett. **16**, 6052 (2016).

[11]  M. Atatüre, D. Englund, N. Vamivakas, S. Y. Lee, and J. Wrachtrup, Nat. Rev. Mater. **3**, 38 (2018).

[12]  M. Widmann, S.-Y. Lee, T. Rendler, N. T. Son, H. Fedder, S. Paik, L.-P. Yang, N. Zhao, S. Yang, I. Booker, and others, Nat. Mater. **14**, 164 (2015).

[13]  A. Gruber, A. Dräbenstedt, C. Tietz, L. Fleury, J. Wrachtrup, and C. Von Borczyskowski, Science. **276**, 2012 (1997).

[14]  M. W. Doherty, N. B. Manson, P. Delaney, F. Jelezko, J. Wrachtrup, and L. C. L. Hollenberg, Phys. Rep. **528**, 1 (2013).

[15] L. Childress and R. Hanson, MRS Bull. **38**, 134 (2013).

[16] A. Katzir, J. T. Suss, A. Zunger, and A. Halperin, Phys. Rev. B **11**, 2370 (1975).

[17] E. Y. Andrei, A. Katzir, and J. T. Suss, Phys. Rev. B **13**, 2831 (1976).

[18] A. W. Moore and L. S. Singer, J. Phys. Chem. Solids **33**, 343 (1972).

[19] A. Sajid, J. R. Reimers, and M. J. Ford, Phys. Rev. B **97**, 064101 (2018).

[20] L. Museur, D. Anglos, J. P. Petitet, J. P. Michel, and A. V. Kanaev, J. Lumin. **127**, 595 (2007).

[21] F. Cataldo and S. Iglesias-Groth, J. Radioanal. Nucl. Chem. **313**, 261 (2017).

[22] W. M. da Silva, H. Ribeiro, T. H. Ferreira, L. O. Ladeira, and E. M. B. Sousa, Phys. E Low-Dimensional Syst. Nanostructures **89**, 177 (2017).

[23] W. Orellana and H. Chacham, Phys. Rev. B **63**, 125205 (2001).

[24] F. Wu, G. Andrew, R. Sundararaman, D. Rocca, and Y. Ping, Phys. Rev. Mater. **1**, 071001 (2017).

[25] G. D. Cheng, Y. G. Zhang, L. Yan, H. F. Huang, Q. Huang, Y. X. Song, Y. Chen, and Z. Tang, Comput. Mater. Sci. **129**, 247 (2017).

[26] M. Abdi, J. P. Chou, A. Gali, and M. B. Plenio, ACS Photonics **5**, 1967 (2018).

[27] B. L. Green, S. Mottishaw, B. G. Breeze, A. M. Edmonds, U. F. S. D'Haenens-Johansson, M. W. Doherty, S. D. Williams, D. J. Twitchen, and M. E. Newton, Phys. Rev. Lett. **119**, 1 (2017).

[28] S. Felton, A. M. Edmonds, M. E. Newton, P. M. Martineau, D. Fisher, and D. J. Twitchen, Phys. Rev. B **77**, 81201 (2008).

[29] B. C. Rose, G. Thiering, A. M. Tyryshkin, A. M. Edmonds, M. L. Markham, A. Gali, S. A. Lyon, and N. P. de Leon, ArXiv:1710.03196 1 (2017).

[30] A. L. Exarhos, D. A. Hopper, R. N. Patel, M. W. Doherty, and L. C. Bassett, ArXiv:1804.09061 (2018).